\documentclass{PoS}
\usepackage[authoryear,square]{natbib}
\usepackage{epstopdf}
\usepackage{color}

\bibpunct{(}{)}{;}{a}{}{,}

\title{Using Tailed Radio Galaxies to Probe the Environment and Magnetic Field of Galaxy Clusters in the SKA Era}

\ShortTitle{Using Extended Radio Galaxies as Environmental Probes}

\author{
\speaker{Melanie Johnston-Hollitt}$^1$\thanks{on behalf of the SKA Cosmic Magnetism Working Group}, 
Siamak Dehghan$^{1}$,
Luke Pratley$^1$
\\ 
$^1$School of Chemical and Physical Sciences, Victoria University of Wellington, P.O. Box 600, Wellington 6140
\\
E-mail: \email{melanie.johnston-hollitt at vuw.ac.nz}
}

\abstract{The morphology of tailed radio galaxies is an invaluable source of environmental information, in which a history of the past interactions in the intra-cluster medium, such as complex galaxy motions and cluster merger shocks, are preserved. In recent years, the use of tailed radio galaxies as environmental probes has gained momentum as a method for galaxy cluster detection, examining the dynamics of individual clusters, measuring the density and velocity flows in the intra-cluster medium, and for probing cluster magnetic fields. To date instrumental limitations in terms of resolution and sensitivity have confined this research to the local $(z < 0.7)$ Universe. The advent of SKA1 surveys however will allow detection of roughly 1,000,000 tailed radio galaxies and their associated galaxy clusters out to redshifts of 2 or more. This is in fact ten times more than the current number of known clusters in the Universe. Additionally between 50,000 and 100,000 tailed radio galaxies will be sufficiently polarized to allow characterization of the magnetic field of their parent cluster. Such a substantial sample of tailed galaxies will provide an invaluable tool not only for detecting clusters, but also for characterizing their intra-cluster medium, magnetic fields and dynamical state as a function of cosmic time. In this chapter we present an analysis of the usability of tailed radio galaxies as tracers of dense environments extrapolated from existing deep radio surveys.}

\FullConference{
Advancing Astrophysics with the Square Kilometre Array\\
June 8-13, 2014\\
Giardini Naxos, Sicily, Italy}

\newcommand{\skipthis}[1]{}

\newcommand{\sd}{deg$^2$}
\newcommand{\as}{$^{\prime\prime}$ }

\begin{document}

\section{Introduction}

It is well known that matter in the Universe is concentrated in a complicated system of twisted filaments of galaxies in the `cosmic web'. At the intersection of these enormous filaments, galaxy clusters are formed and evolve. Understanding how these massive systems are created and evolve is crucial to our perception of the large-scale structure of the Universe. In recent years various techniques and advanced instruments have been established or enhanced to detect and characterize filaments, groups, and clusters. These methods are predominantly undertaken in parts of the spectrum other than the radio, and include X-ray \citep{hbg82} and Sunyaev-Zeldovich \citep{sz80} detections, and statistical data analysis and clustering examinations of spectroscopic and photometric redshift surveys \citep{hg82,dj14}. However, recent radio wavelength studies of groups and clusters have shown that some types of extended radio sources trace high-density regions of the Universe, and moreover, they provide valuable information on properties of large-scale structures of the Universe. This use of radio sources to probe large-scale structure and understand environmental conditions is set to increase in the Square Kilometre Array (SKA) era (\citealp{nab13,cbb14,fds14,gtb14,nbb14}).

In this context, the community's attention has recently focused on a class of radio galaxies known as Bent-Tailed sources (BTs) as probes to both locate high density environments \citep{bgh01,mss10,wb11} and to study the conditions in these regions \citep{pj11}, including their local magnetic field (\citealp{eo02,pjd13}). BTs as a class include radio galaxies with asymmetric radio structure in which lobes or plumes are not lined up with the galaxy itself. A subdivision of such sources with a morphology demonstrating that the jets are bent back through large angles are known as Head-Tail (HT) or Narrow-Angle-Tail (NAT) radio galaxies. HTs, which are almost exclusively found in the vicinity of galaxy clusters, are classified as Fanaroff-Riley class I radio sources \citep{fr74}. The curved radio structure of HTs is believed to be induced by ram pressure due to the relative movement of the host galaxy within the Intra-Cluster Medium (ICM; \citealp{gg72}). However, \citet{cm75} suggested buoyancy forces due to density perturbations in the ICM as an alternative mechanism responsible for the curvature formation in HTs and other BT radio sources. The coexistence of BTs and clusters in the local $(z \leq 0.1)$ Universe, is at a rate of 1-2 per cluster \citep{mjs09}, and there is growing evidence that such associations carry on as far as redshift 2, at the limits of cluster detection \citep{djf14}. Thus, the correlation between BTs and galaxy clusters provides a powerful tool to locate the high density regions of the Universe, and additionally, investigate the physical characteristics of rich environments.

The SKA and its precursors will detect BTs in large quantities enabling a range of science goals to be met. Here we review the science that can be derived from BTs and discuss their expected detection numbers with SKA surveys.

\begin{figure*}
\centering
{\vspace{-0cm}\hspace*{-0cm}\includegraphics[width=2.7in, trim= 0 0 0 0, clip=true]{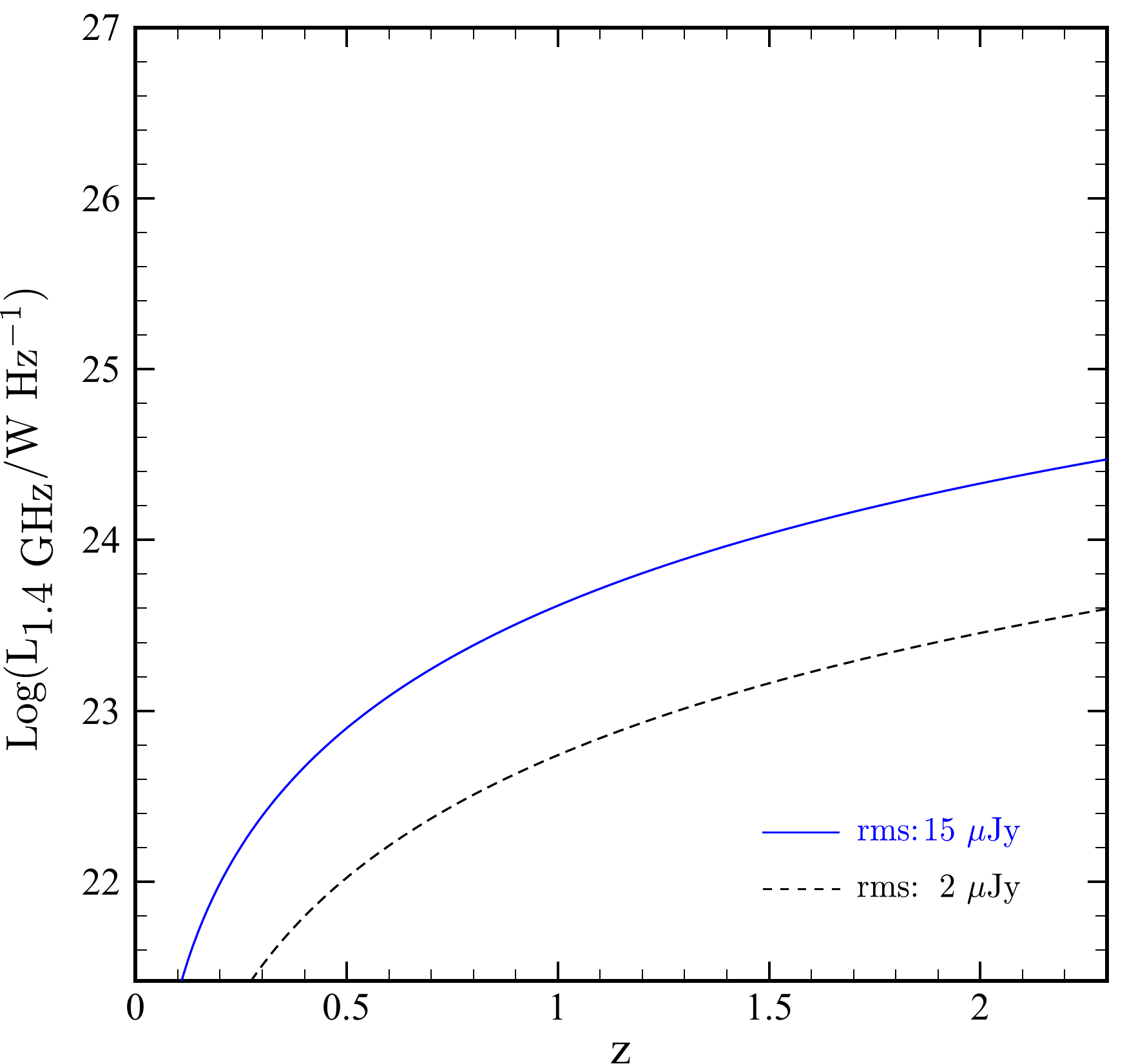}}
{\hspace*{-0cm}\includegraphics[width=2.7in, trim= 0 0 0 0, clip=true]{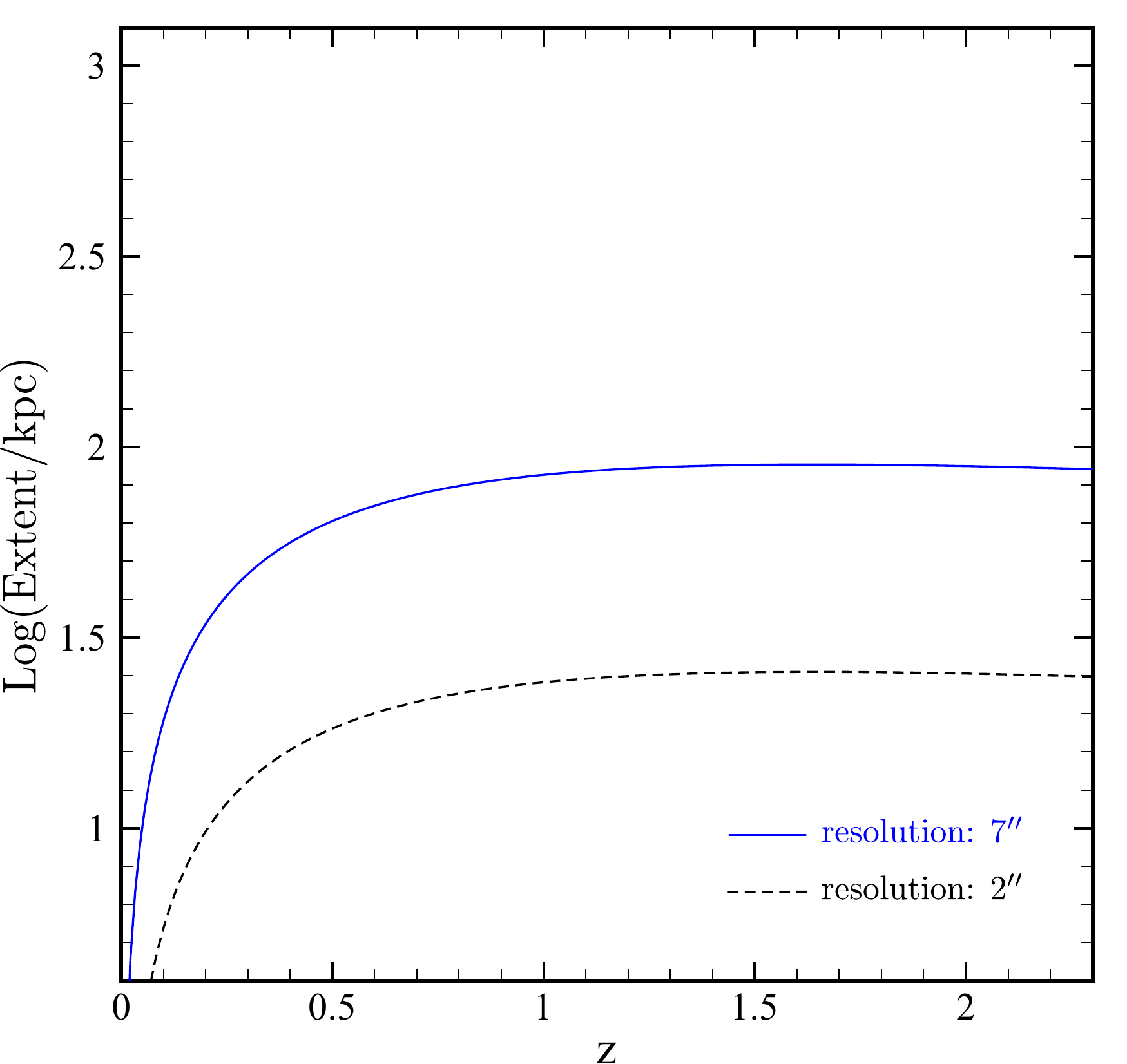}}
\vspace{-0cm}\caption{Left panel: The blue line and black dashed lines represent the 6 sigma source detection limit of the ATLAS and SKA1 surveys with the average sensitivity of 15 and 2 $\mu$Jy, respectively. Right panel: The blue line and black dashed line represent the detection limit of the sources with an extent of 1.5 beams in ATLAS and SKA1 surveys at 7 \& 2 arcsec resolution, respectively.}
\label{fig:limit}
\end{figure*}

\section{Detecting Tailed Radio Galaxies in the SKA Era}

A series of tiered surveys with increasing sensitivity over smaller and smaller areas are planned to be undertaken with the Square Kilometre Array during phase 1 (SKA1; \citealp{nab13,nbb14}). This approach, known as a `wedding cake' survey strategy, will commence with an `all-sky' survey (31,000 \sd) from 1-2 GHz with 2\as resolution and 2 $\mu$Jy sensitivity. Such a survey can be expected to detect numerous BT galaxies and therefore their associated clusters.

In order to assess the number of BT sources likely to be detected with SKA surveys, we can look to detection rates in deep surveys of SKA-level sensitivity carried out on smaller areas. \citet{djf14} performed a comprehensive radio source investigation in the 4 \sd area of the Australia Telescope Large Area Survey of the Chandra Deep Field-South (ATLAS-CDFS) field, in order to evaluate the ability of SKA1 surveys to identify BT radio sources. The ATLAS-CDFS which has been observed with the Australia Telescope Compact Array at 1.4 GHz (\citealp{naa06}; T.~M.~O.~Franzen et al., in preparation), has about 17\as$\times$ 7\as resolution and a depth of 15 $\mu$Jy/beam. In addition, the ATLAS-CDFS image is partially covered by a 2\as resolution 1.4 GHz Very Large Array (VLA) image of the Extended CDFS (ECDFS) with an rms noise down to 10 $\mu$Jy/beam \citep{mbf13} which is well matched to the resolution and sensitivity of the planned SKA1 surveys (Norris et al.\ 2014). It is worthwhile to note that while the expected capabilities of the SKA1 all-sky survey are similar in resolution to the VLA-ECDFS image, the SKA1 survey will be marginally more sensitive (see Figure \ref{fig:limit}). \textcolor{black}{As Figure} \ref{fig:limit} \textcolor{black}{demonstrates the SKA1 surveys will detect BT sources of lower power and smaller physical extent then are currently accessible with present wide-field surveys.}

Out of a total of $\sim 3000$ radio sources in the field, \citet{djf14} found 56 extended and diffuse radio sources, consisting of 45 BTs, a radio relic candidate, and several complex diffuse sources which could not be unambiguously categorized. Based on extrapolating from these detections, forthcoming all-sky 1.4 GHz continuum surveys with a resolution and sensitivity comparable to the ATLAS survey undertaken on pathfinder instruments such as the EMU survey \citep{nha11} on ASKAP, will identify over 590,000 extended and diffuse low-surface-brightness radio sources, including at least 470,000 BT radio galaxies and the SKA1 level surveys will double this number to of order $\sim$ 1 million BT sources. Statistical studies of such large samples of tailed radio sources will boost our understanding of the evolutionary origins of BTs and the mutual dependency between these intriguing radio sources and their habitat.

The postulated association between BT radio sources and rich environments suggests that the SKA1 will detect more than 1 million clusters and groups, which is more than twenty times the number of clusters identified at present. This, in fact, surpasses the number of clusters expected to be found with future X-ray telescopes such as eROSITA \citep{mpb12} making SKA a powerful new tool for cluster detection. Additionally, assuming that these clusters and groups are found by both eROSITA and SKA1 surveys, there will be an exceptional opportunity to investigate the multi-wavelength characteristics of high-density regions of the Universe up to a redshift of 2.

\section{Tailed Radio Galaxies as Environmental Probes}

BT radio galaxies represent a wide range of morphologies that are believed to be primarily generated by environmental effects \citep{b98}. Two significant mechanisms are ram pressure and buoyancy forces induced on lobes of tailed radio galaxies. These forces are believed to be created by either relative motion of the host galaxy with respect to the ICM, or considerable deviations in the density of the medium. In particular, radio galaxies falling into a cluster's gravitational well, or sources within a turbulent weather system of clusters in a merger process, are thought to develop curved lobes and jets. On more local scales, the structure of jets in BTs may be notably altered by gravitational influences of satellite and nearby galaxies via bound and unbound orbital motions, near-miss passes, and precession of the accretion disk and jets. As a result BTs have been used with increasing success to probe conditions in their local environment. For example the particular morphology of tailed radio galaxies, if well resolved, can be used to investigate the dynamical state of their parent clusters such as in Perseus \citep{pj11}, A3135 \citep{pjd13} and A3266 (\citealp{d14}; Miller et al., in preparation) where evidence for shocks, cluster winds and the merger direction, respectively, were all provided by the BT galaxies. BTs have also successfully been used to measure the density and velocity flows in the intra-cluster medium (\citealp{fcw08}; \citealp{dbc11}), and for measuring cluster magnetic fields which will be discussed in the next section. Crucially though, all such studies have been undertaken only on individual objects and the wealth of information to be gain from undertaking statistical studies of this type have yet to be realised, though recent steps in this direction have commenced with existing surveys such as FIRST \citep{wb11}.

\begin{figure*}
{\vspace{-0.3cm}\hspace{0.15cm}\includegraphics[width=3.2in, trim= 25 0 40 0, clip=true]{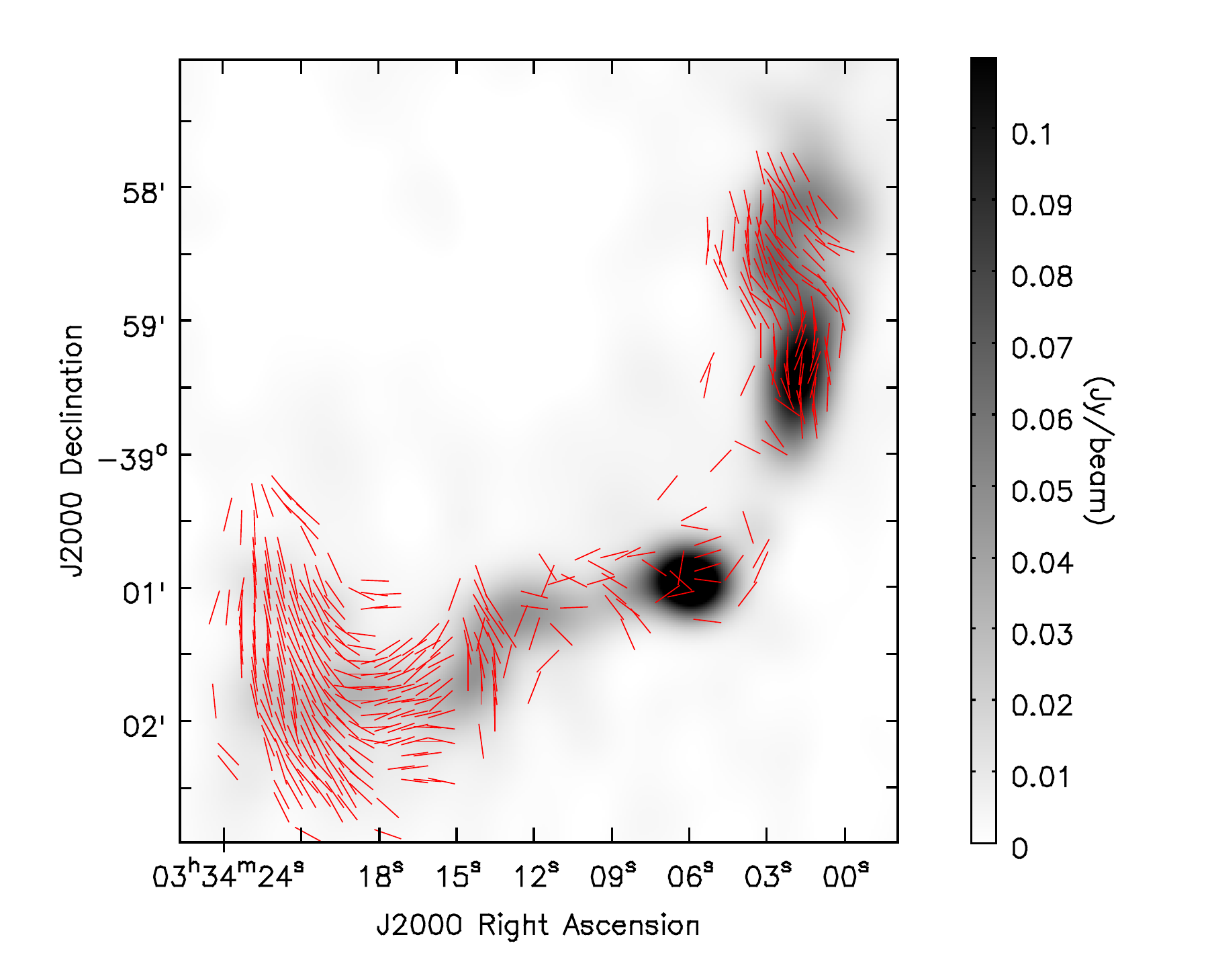}}
{\includegraphics[width=3.2in, trim= 307 0 10 0, clip=true]{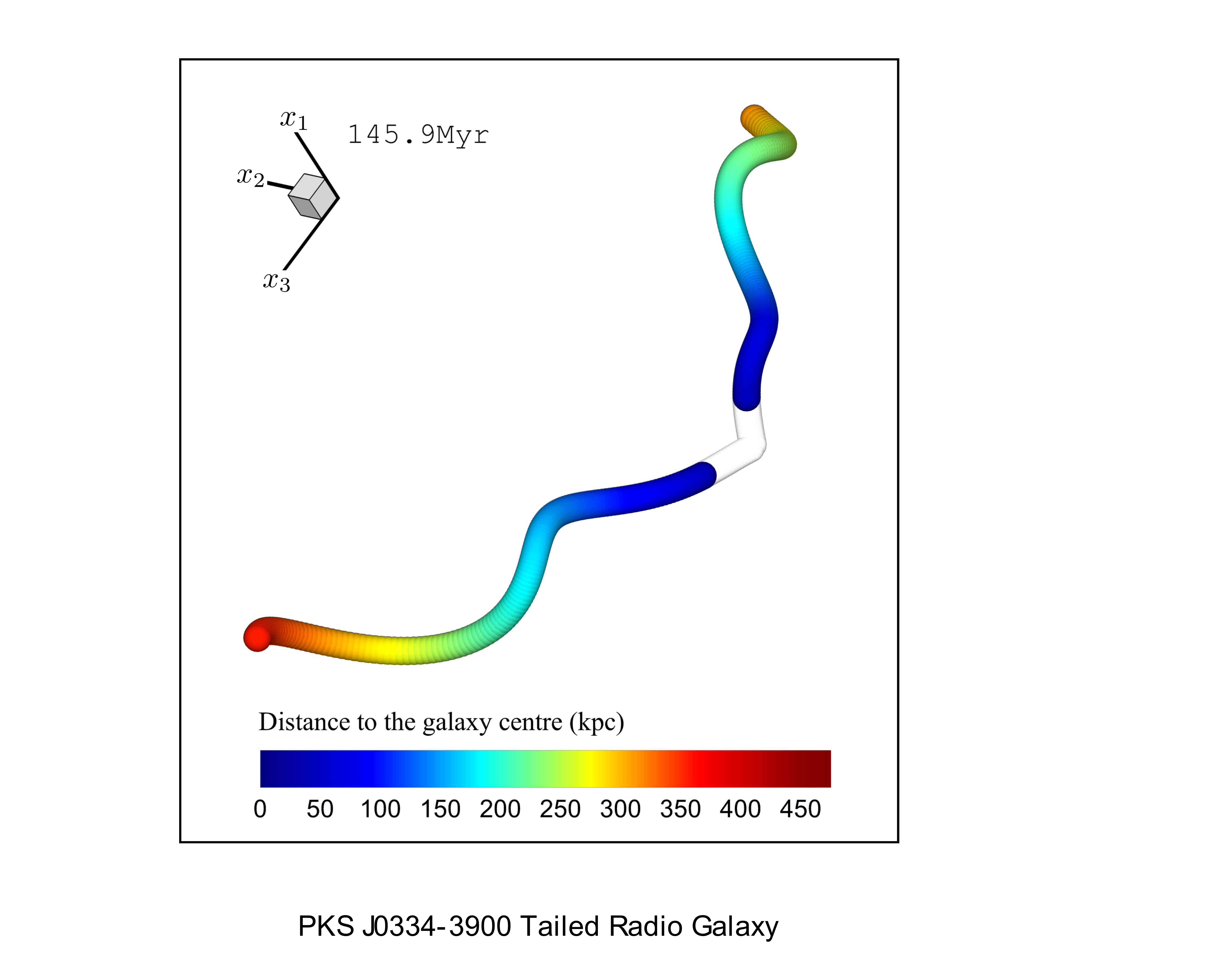}}
\vspace{-0.7cm}\caption{Left panel: Red segments represent the direction of the magnetic component of the polarized emission overlaid on the 1.4 GHz intensity image of the tailed radio galaxy PKS J0334-3900 at the centre of Abell 3135. Right panel: The simulated radio structure of the BT shown in the left panel. For further information see \citet{pjd13}.}
\label{fig:rm}
\end{figure*}

An illustration of the wealth of information to be obtained by BTs is presented in \citet{pjd13} where a simple mechanical model was developed that simulates the most common observed BT shapes, including V-, M-, S-, and X-shaped radio galaxies, based on likely movement of the host galaxy under different environmental effects. This was then applied to simulate a peculiar tailed radio galaxy in the Abell 3135 galaxy cluster (see Figure \ref{fig:rm}). This study was able to disentangle various effects at work and showed that a cluster wind combined with both orbital and precessional motions must be present in order to create the observed morphology of this BT radio source. This simple, yet powerful framework applied to a statistical sample of BTs in clusters could elucidate such questions as: 
\begin{itemize}
\item how often do we find cluster winds and how strong are they?
\item Do BTs always require a companion galaxy and does this account for their association with clusters and dense groups?
\item How often do AGN in BTs precess and can this explain X-shaped radio galaxies alone without having to require black hole - black hole collisions?
\end{itemize} 

The latter question having important implications for recent predictions of the rate of black hole in-spirals and hence generation of gravitational waves expected to be found with next generation detectors such as Advanced LIGO and LISA.


\subsection{Using tailed radio galaxies to constrain cluster magnetic fields}

While the morphology of tailed radio galaxies provides a probe of the cluster weather, their linearly polarized synchrotron emission provides a probe of the cluster magnetic field. The polarized emission from the jets of the tailed radio galaxy (see Figure 2 for an example) passes through the intra-cluster medium and magnetic field, causing the linear polarization angle to rotate due to Faraday rotation. Knowing the path length (distance) over which the Faraday rotation occurs and the electron density of the intra-cluster medium (available from X-ray observations), it is possible to constrain the strength of the intra-cluster magnetic field.

Cluster magnetic fields are not well understood because they are difficult to measure, however, they are believed to play an important role in the cluster environment. While several authors have focused on the way cluster magnetic fields can be studied in the radio, including the 2004 SKA Science Case chapter by \citet{fj04}, such works tend to concentrate on the detection of diffuse cluster emission such as relics or statistical studies of background galaxies to probe the intra-cluster magnetic fields. Whilst such techniques remain popular and are the subject of several chapters in the new SKA Science Case (\citealp{gmx14,bvb14}), observations of tailed radio galaxies in the last two decades have demonstrated that resolved BTs have a valuable role in constraining the coherence length of magnetic fields in clusters on smaller scales than other methods, e.g. Coma \citep{fdg95,bfm10}, A119 \citep{fdg99,mgf04}, A2382 \citep{gmg08} and 3C129 \citep{tga01}. 

In one of the first works to utilise resolved BTs, \citet{eo02} argued that the Faraday rotation across the tailed radio galaxies in A400 and A2634 is due to the Faraday screen of the intra-cluster medium and its magnetic field, rather than a skin of the radio source. They went on to show that the galaxy clusters have magnetic fields ordered on the scale $\sim 10-20$ kpc, and the geometry of the rotation measure appears to be determined by the intra-cluster medium with magnetic field strengths estimated to be $\sim$1-2 $\mu$G. Similarly, \citet{j03} argued that the coherence length of BTs resolved in A3667 was of order of $10-20$ kpc. Furthermore, \citet{ve03} used the Faraday screen in the foreground of tailed radio galaxies in A400, A2634 and Hydra to measure the power spectrum of the intra-cluster magnetic fields finding a range of magnetic field autocorrelation lengths on small scales. \citet{gmf06} investigated using simulations of magnetic fields in A2255 to reproduce the Faraday screens observed from BTs, thus constraining the magnetic field structure. Following the method of Rotation Measure Synthesis \citep{bb05}, \citet{pbb11} first explored the use of BTs as resolved screens at different depths within A2255 using the polarization properties as a tool to unveil the position of the radio sources within the cluster. In the study of \citet{pjd13}, the morphology of a tailed radio galaxy is used to estimate the distance at which Faraday rotation occurs within the cluster A3135. Whereas in \citet{pbb11} the location of the resolved BTs within A2255 along the line of site was not precisely known, in \citet{pjd13} there was a wealth of multiwavelength data to understand the location of the BT jets within the cluster. As mentioned above, the BT was modelled so as to constrain several parameters required to generate the observed morphology, including detailed modelling of the jet locations/projection. From this model, it was possible to estimate the distance between the two jets of the radio galaxy along the line of sight. The difference in average Faraday rotation of each jet could then be explained by the separation of the jets along the line of slight, allowing one to estimate the magnetic field over the separation (160 kpc) giving an estimate of the field over small scales within the cluster.

As with the improved statistics SKA1 is expected to yield for the use of extragalactic background source rotation measures (\citealp{mjh14}; \citealp{bvb14}), the use of resolved tailed radio galaxies as probes of the magnetic fields in clusters will increase by several orders of magnitude. While detailed modelling is required to determine the expected number of sources with sufficient polarized emission for such studies, we may make a back of the envelope estimate by extrapolating from current surveys. This would suggest that approximately 5 to 10 per cent of the 1 million BTs to be detected with SKA1 should have sufficient polarization detected for such studies, giving between 50,000 and 100,000 resolved BTs in clusters for use as probes of the cluster magnetic field. Combined with the other sources of the expected 40 million extragalactic polarized sources detected as part of the RM survey experiment, the BTs embedded within clusters will form a powerful complementary dataset to disentangle cluster magnetic fields and perform tomographic mapping of magnetic fields in galaxy clusters. This will be analogous to the science case for combining background extragalactic RMs with pulsars and diffuse polarization to probe the magnetic field of the Milky Way, and is set to drastically improve our knowledge of the extent of cluster magnetic fields and their structure on small scales. Importantly, with such a large sample we will be able to probe the evolution of cluster magnetic fields as a function of time, potentially evaluating how the field geometry and strength change, this will give vital clues to understand both the evolution and origin of magnetic fields in the Universe.

\section{Impact of SKA Design}

Successful searches for BT galaxies require both sensitivity and resolution. Looking forward to SKA2 we can expect to detect even greater numbers of BTs, potentially at even higher redshifts, signalling the first significant movement of objects in dense atmospheres, potentially in the very earliest of clusters.

Potential changes of design to the SKA that might affect the use of BT galaxies to detect and probe the environment of galaxy clusters include both a loss of resolution, which would reduce the redshift range over which such sources can be reliably detected if their physical sizes are small, and a loss of sensitivity which would reduce the chances of detecting faint tailed sources.

Recent studies have shown that BT sources can be as small as $\sim$50 kpc \citep{djf14}, which if located in a very distant cluster would require at least 1" resolution to unambiguously detect. The impact of sensitivity to this type of science can best be illustrated through the number of BTs detected in the original ATLAS-CDFS field (\citealp{mss10}; \citealp{djm11}) as compared to those found in the 3rd data release \citep{djf14} in which the resolution was identical but the sensitivity doubled increasing the number of BTs by a factor of 5. This suggests that using, say, only of 50$\%$ of the currently planned number of antennas would reduce the total number of BTs detected to 200,000 in terms of sensitivity alone. If the reduction also resulted in a decrease in baseline length, this number will reduce even further due to inability to resolve sources. Of the detected sources with such an array, only 10,000 - 20,000 would then be suitable for polarimetric studies of cluster magnetic fields. These numbers are still large and thus, so long as the resolution were not substantially compromised, a full sky survey with a 50$\%$ SKA1 could still advance this type of science considerably.

\section{Summary}

Future all-sky radio continuum surveys via instruments such as SKA1 will provide more sensitive and yet observationally affordable data to explore a wide range of science goals including those associated with the detection and characterisation of the environment of up to 1 million galaxy clusters using BT galaxies. Combining these data with that of future X-ray missions will provide an exceptional dataset to investigate the dynamical history of galaxy clusters. In addition, between 50,000 and 100,000 detected BTs will be resolved and of sufficient polarization to be used as probes of cluster magnetic fields. Combined with background polarized sources from the RM Survey, the embedded cluster BTs will allow the first tomographic view of cluster magnetic fields as a function of cosmic time. From this we may finally answer questions about the origin and evolution of cluster magnetic fields and how they influence cluster dynamics.

\bibliographystyle{apj}
\bibliography{BB}

\end{document}